\documentstyle[11pt]{article}
\begin{document}

 \newcommand \be {\begin{equation}}
\newcommand \ee {\end{equation}}
 \newcommand \ba {\begin{eqnarray}}
\newcommand \ea {\end{eqnarray}}

\newcommand{\siml}{\stackrel{<}{\sim}}
\def \p {\vec \phi({\bf x})}
\def \dk { {\int  \frac{d^D{{{\bf k}}}{(2\pi)^D}} }}
\def \(({\left(}
\def \)){\right)}
\def \[[{\left[}
\def \]]{\right]}

\title{\bf THE LARGE SCALE ENERGY LANDSCAPE OF RANDOMLY PINNED OBJECTS}

\vskip 3 true cm

\author{Leon Balents$^1$, Jean-Philippe Bouchaud$^2$ and Marc M\'ezard$^3$}

\date{\it $^1$  Institute of Theoretical Physics, \\ Santa Barbara, CA 93106-4030   , USA \\
$^2$ Service de Physique de l'\'Etat Condens\'e,
 Centre d'\'etudes de Saclay, \\ Orme des Merisiers, 
91191 Gif-sur-Yvette Cedex, France \\ 
$^3$ Laboratoire de Physique Th\'eorique de l'Ecole Normale Sup\'erieure
 \footnote {Unit\'e propre du CNRS,  associ\'ee
 \`a\ l'Ecole
 Normale Sup\'erieure et \`a\ l'Universit\'e de Paris Sud} , \\
24 rue
 Lhomond, 75231 Paris Cedex 05, France }


\maketitle

\begin{abstract}
{We discuss the large scale effective potential for elastic objects
(manifolds) in the presence of a random pinning potential, from the
point of view of the Functional Renormalisation Group (FRG) and of the
replica method. Both approaches suggest that the energy landscape at
large scales is a succession of parabolic wells of random depth,
matching on singular points where the effective force is
discontinuous.  These parabolas are themselves subdivided into smaller
parabolas, corresponding to the motion of smaller length scales, in a
hierarchical manner. Consequences for the {\it dynamics} of these
pinned objects are underlined.}
\end{abstract}

\vskip 0.5cm

LPTENS preprint 96/06.

\vskip 0.5cm

\noindent Electronic addresses : balents@itp.ucsb.edu, 
bouchaud@amoco.saclay.cea.fr, mezard@physique.ens.fr

\newpage

\section{Introduction}
The physics of elastic objects pinned by random impurities is
certainly one of the most topical current themes of statistical
mechanics. The problem is of fundamental importance both from a
theoretical point of view (many of the specific difficulties common to
disordered systems are at stake) and for applications: the pinning of flux lines in
superconductors \cite{FFH,Vortex,Kardar}, of dislocations, of domain walls in magnets, or of
charge density waves \cite{FLR,FXY}, controls in a crucial way the
properties of these materials. Interestingly, this problem is also
intimately connected to surface \cite{HH} and crack growth
\cite{cracks} and to turbulence \cite{BMP}.

Two different general approaches have been proposed to describe the
{\it statics} of these pinned manifolds, for which perturbation theory badly fails. The first one is the
`functional renormalisation group' (FRG) which aims at constructing
the correlation function for the effective pinning potential acting on
long wavelengths using renormalisation group (RG) ideas
\cite{F,BF}. The second is the variational replica method which
combines a Gaussian trial Hamiltonian with `replica symmetry breaking'
to obtain results in the low temperature, strongly pinned phase
\cite{MP,BMY,LDG}. Although many of the results of these two approaches
actually turn out to be similar \cite{P,M,MP,FH2,HF,LDG}, the feeling that
the link between them is missing is rather widespread, reflecting the
fact that our present general understanding of disordered system is
still incomplete.

The aim of this letter is to unveil precise connections between
these two (sometimes presented as conflicting \cite{FH1,FH2,BF})
theories. We show that both formalisms are indeed struggling to
describe an awkward reality: the effective, long wavelength pinning
potential has the shape drawn in Fig.1. It is a succession of
parabolic wells of random depth, matching on singular points where the
effective force (i.e. the derivative of the potential) is
discontinuous. These discontinuities induce a
singularity in the effective potential correlation function, and are
encoded in the replica language by the RSB. The replica calculation
furthermore provides an explicit construction of this effective
(random) potential, and hence, in turn, information on the statistics
of -- say -- the depth of the potential minima.  The replica
calculation might also shed light on the domain of validity of the
FRG, by making more explicit the assumptions on which the latter
relies.

Apart from the satisfying possibility of reconciling two rather
different microscopic methods, we believe that our construction is
very useful to understand the {\it dynamics} of such objects. For
example, their relaxation can be analyzed in terms of hops between
the different minima (`traps'), corresponding to metastable long
wavelength configurations. The statistics of barrier heights control
the trapping time distribution, and hence the low frequency response
and its possible aging behaviour \cite{Bou92,DFI}.  Another interesting
situation is the zero temperature depinning transition induced by
an external driving field, which has recently been investigated, again
using RG ideas for expanding around a mean-field limit
\cite{Natt,NF,EK}. However, the results depend on the form of the
pinning potential in this mean-field limit. The correct form was
surmised by Narayan and Fisher \cite{NF} to be the `scalloped'
potential of Fig 1. Our calculation, to some extent, confirms their
intuition.

The model we consider is the (by now standard) Hamiltonian describing
pinned elastic manifolds: 
\be {\cal H}(\{\p\})= \int d^D{{\bf x}}
\quad \[[ \frac{c}{2}\((\frac{\partial \p}{\partial {{\bf x}}}\))^2 +
V_0({{\bf x}},\p) \]], \ee 
where ${{\bf x}}$ is a $D$-dimensional
vector labelling the internal coordinates of the object, and $\p$ an
$N$-dimensional vector giving the position in physical space of the
point labelled ${{\bf x}}$. Various values of $D$ and $N$ actually
correspond to interesting physical situations. For example, $D=3$,
$N=2$ describes the elastic deformation of a vortex lattice (after a suitable anisotropic generalisation of Eq. (1)), $D=2$,
$N=1$ describes the problem of domain walls pinned by impurities in 3
dimensional space, while $D=1$ corresponds to the well-known directed
polymer (or single flux line) in a $N+1$ dimensional space.  The elastic
modulus $c$ measures the difficulty of distorting the structure, and
$V_0({{\bf x}},\p)$ is a random pinning potential, which we shall
choose to be Gaussian with a short range correlation function: 
\be
{\overline{V_0({{\bf x}},\vec \phi) V_0({{\bf x}}',\vec \phi')}}_c =
{ N W } \delta^D({{\bf x}}-{{\bf x}}') R_0\((\frac{(\vec \phi-\vec
\phi')^2} {N}\)), \label{R0} \ee 
where $W$ measures the
strength of the pinning potential.  In the following, we shall choose for convenience $R_0(y)=\exp(-\frac{y}{2 \Delta^2})$ where $\Delta$ is the correlation length of the random potential.

One aim of the theory is to understand how the microscopic pinning
potential will affect the elastic manifold on long length scales,
relevant for macroscopic measurements. In other words, one would like
to construct the {\it effective} pinning potential seen by a low
wavevector mode of the structure, after thermalizing the modes with
shorter length scales.  Both the FRG and the replica approach propose
an approximate construction of this effective potential which we now
discuss and relate.

\section{The Functional Renormalisation Group.}
In the spirit of the momentum shell renormalisation group, the FRG
method consists in writing down a recursion relation for the
correlation function of the potential acting on `slow' modes $\vec
\phi_<$, after `fast' modes $\vec \phi_>$ (corresponding to
wavevectors in the high-momentum shell $[\Lambda/b,\Lambda]$) have
been integrated out using perturbation theory. This procedure has been
addressed in considerable detail in Ref.~\cite{BF}, we present only a
brief description of the calculations.  At zero temperature the
renormalized Hamiltonian is defined by $H_{\rm R}[\vec \phi_<] =
\int_{{\bf x} } \frac{1}{2} |\nabla\vec \phi_<|^2 + V_{\rm R}[\vec \phi_<]$ and
\be {\cal V}_{\rm R}[\vec \phi_<] = \min_{\vec \phi_>} \int_{{\bf x} }
\bigg\{ {1 \over 2} |\nabla\vec \phi_>|^2 + V(\vec \phi_< + \vec
\phi_>,{{\bf x} })\bigg\}, \label{mindef} \ee 
where the original
field $\vec \phi = \vec \phi_< + \vec \phi_>$ has been split into low
($\vec \phi_<$) and high ($\vec \phi_>$) momentum components.  The
renormalized Hamiltonian $H_{\rm R}$ thus describes the long-distance
physics of modes with momenta $k < \Lambda/b$, where $\Lambda$ is the
original short-scale cutoff, and the rescaling factor $b>1$.  The FRG
proceeds to determine the minimum in Eq. (\ref{mindef}) perturbatively
in $\vec \phi_>$.  The extremal condition may be expanded in $\vec
\phi_>$ as \ba -\nabla^2 \phi_>^i & = & - \partial_i V(\vec
\phi_<+\vec \phi_>,{{\bf x} }) \nonumber \\ & \approx & -\partial_i
V(\vec \phi_<,{{\bf x} }) - \partial_i\partial_j V(\vec
\phi_<,{{\bf x} }) \phi_>^j. \label{expand} \ea 
where $\partial_i \equiv \frac{\partial}{\partial \phi^i}$. Defining the Fourier
transform $\tilde{V}^{ij\cdots}_{\bf k} = \partial_i \partial_j
\cdots \int_{{\bf x} } V(\vec \phi_<,{{\bf x} })e^{-i{\bf k}\cdot{{\bf x} }}$, the approximate solution is 
\be \Lambda^2
\phi_{>,{\bf k}}^i \approx - \tilde{V}^i_{\bf k} + \Lambda^{-2}
\int_{{\bf k}'} \tilde{V}^{ij}_{{\bf k}-{\bf k}'} \tilde{V}^j_{{\bf k}'}. \label{phiapprox} \ee 
Inserting this solution into the energy
(Eq. (\ref{mindef})) gives 
\be {\cal V}_{\rm R} = \tilde{V}_0 - {1 \over
{2\Lambda^2}}\int_{\bf k}^> \tilde{V}^i_{\bf k}\tilde{V}^i_{-{\bf k}} + {1 \over {2\Lambda^4}} \int_{{\bf pp }'}^> \tilde{V}^{ij}_{{\bf k}+{\bf k}'} \tilde{V}^i_{-{\bf k}} \tilde{V}^j_{-{\bf k}'}. \ee
where $\int^>$ is restricted to the high-momentum shell. 
If $\vec \phi_<({{\bf x} })$ is constant over regions of size $\ell$,
this can be rewritten as an integral of a {\sl local} potential, up to
small errors of order $1/\ell$: ${\cal V}_R(\vec \phi_<) \simeq \int d{\bf x} V_R(\vec \phi_<,{\bf x})$.  Thus, in the long wavelength limit,
the renormalized Hamiltonian is well-described simply by a
renormalized potential.  Its connected correlations can be calculated from the
expression 
\be \overline{V_{\rm R}(\vec \phi,{\bf x}) V_{\rm R}(\vec
\phi',{\bf x}')}_{\rm C} = R_{\rm R}\((\frac{[\vec \phi-\vec \phi']^2}{N}\)) \delta({\bf x}-{\bf x}'). \ee 
Assuming that the statistics of the effective potential
remains Gaussian, one finds within this first order perturbation
theory: 
\be R_{\rm R}(y) = R(y) + {{dl} \over {8\pi^2}}\bigg[ {1
\over 2}\partial_i\partial_j R\partial_i\partial_j R -
\partial_i\partial_j R \partial_i\partial_j R(0)\bigg]
\label{modefinal} \ee
in $D=4$, where $b = e^{dl}$ and $dl$ is infinitesimal.

Eq. (\ref{modefinal}) is the final result of the mode elimination.
The search for fixed points requires the additional step of a
rescaling transformation, which restores the original value of the
cutoff $\Lambda$.  Performing this rescaling via ${{\bf x}}
\rightarrow b{{\bf x}}$ and $\phi \rightarrow b^\zeta \phi$ results
in the full RG equation for the correlator 
\be \partial_l R =
(\epsilon-4\zeta)R + \zeta \phi^i\partial_i R + {1 \over
{8\pi^2}}\bigg[ {1 \over 2}\partial_i\partial_j R\partial_i\partial_j
R - \partial_i\partial_j R \partial_i\partial_j
R(0)\bigg]. \label{Rflow} \ee 
Iteration of this equation from the
`initial' condition $R(y)=R_0(y)$ converges towards the {\it fixed
point} $R^*(y)$, describing the long wavelength properties, which has
the singular small $y$ expansion \cite{BF} 
\be R^*(y) -
R^*(0) = \epsilon y [ a_1  - a_{3/2} \sqrt{y}] +
..., \label{R*} \ee 
where $\epsilon = 4-D$ is the small parameter
justifying the use of perturbation theory. Another way of stating this
result is in term of the effective {\it force} $f$ acting on the
manifold, defined as minus the derivative of the effective potential
with respect to $\phi$. The force correlation function then behaves as
\be \overline{ [f^*(\phi)-f^*(\phi')]^2 } = 12 \epsilon a_{3/2} |\phi
- \phi'|. \label{f*} \ee 
Together with the assumption of Gaussian
statistics, this suggests that the effective force acting on the
manifold behaves, for $N=1$, as a {\it random walk} in $\phi$
space. This picture was advocated in \cite{BF}, and was actually used
to argue that the next correction in $\epsilon$ to would be of order
$\epsilon^{\frac{3}{2}}$.

\section{The replica approach.}

The replica approach is, in some sense, more ambitious, since it
 provides an explicit probabilistic construction of the effective
 disordered potential seen by the manifold. On the other hand, the
 method can only be controlled in the $N \to \infty$ limit, where a
 Gaussian variational Hamiltonian becomes exact \cite{unsurN}.  Let us
 however stress right away that a Gaussian Hamiltonian in replica
 space {\it does not} mean that the actual effective potential which
 we wish to characterize has Gaussian statistics. As we shall indeed
 show below, this is not at all the case.

Let us sketch first how the correlation function $R(y)$ can be
calculated with replicas and compared with the FRG. (More details can
be found in \cite{MP,BMY,BMP}). The average free-energy $F
=-\frac{1}{
\beta}\overline{\ln Z} \equiv -\frac{1}{\beta}\overline{\ln
\int {\cal D}\phi \exp [-\beta {\cal H}]}$ is computed as usual as the
`zero replica' limit $\ln Z = \lim_{n \to 0} {Z^n -1 \over n}$.  The
average of $Z^n$ can be seen as the partition function of the
following $n-$replica Hamiltonian: 
\be {\cal H}_n = {c \over 2}
\sum_{a=1}^n \int d^D{{\bf x}} \left( {d\vec \phi^a \over
d{{\bf x}}} \right)^2 -{W N \over 2} \sum_{a,b} \int d^D{{\bf x}}
\exp \left[-{(\vec \phi^a({{\bf x}})-\vec \phi^b({{\bf x}}))^2 \over
2N\Delta^2}\right], \ee 
where an effective attraction between replicas
has emerged from the disorder average. The idea is to treat this
interaction using the trial Hamiltonian \cite{Orland} 
\be {\cal H}_v \equiv {1 \over 2} \sum_{a,b} \sum_{{{\bf k}}} \ \vec
\varphi^a(-{{\bf k}}) G^{-1}_{ab}({{\bf k}}) \vec
\varphi^b({{\bf k}}), \ee 
where $\varphi^a({{\bf k}}) \equiv
L^{-\frac{D}{2}} \int d^D {{\bf x}} \ \phi^a({{\bf x}}) e^{-i
{{\bf k}} \cdot {{\bf x}}}$, and $L$ is the `linear' size of the
manifold.

The trial free-energy obtained with ${\cal H}_v$ depends on $G_{ab}$
and reads ${\cal F}_v[G] = <{\cal H}_n>_v - {N \over 2
\beta} Tr \ln
G$; the optimal matrix $G$ is then determined by minimizing ${\cal
F}_v[G]$, which leads to a set of self-consistent equations for
$G_{ab}$. The point now is that the structure of $G_{ab}$ in replica
space can be non trivial in the limit $n \to 0$, corresponding to
`replica symmetry breaking'. The physical meaning of this procedure
has already been described in detail in \cite{MP,MPV,BMP}, and we
shall come back to it later. Before describing the solution to these
self-consistent equations in the regime $D \siml 4$, one should
clarify first in what sense the replica calculation allows one to
characterise the large scale pinning potential. Since the trial
Hamiltonian is factorized over Fourier modes, one can isolate a
particular, very slow mode ${{\bf k}}_0 \to 0$. The effective force
acting on $\vec \varphi_0 \equiv \vec \varphi({{\bf k}}_0)$ is
$f^\mu_\Omega(\vec \varphi_0) = -\frac{1}{\beta} {\partial \over
\partial \varphi_0^\mu} \ln {\cal P}_\Omega(\vec \varphi_0)$, where
${\cal P}_\Omega(\vec \varphi_0)$ is the probability to observe $\vec
\varphi_0$ for a given realisation of the random pinning potential
$\Omega$. It is thus clear that in order to compute, say, the
correlation function of $\vec f$, one should study the 
object: 
\be \overline {f^\mu_\Omega(\vec \varphi_0)f^\nu_\Omega(\vec
\varphi_0')}= \lim_{n \to 0} {4 \over n^2} \frac{\partial^2 }
{\partial \varphi_0^\mu \partial \varphi_0'^\nu} \overline {[{\cal
P}_\Omega(\vec \varphi_0)]^{\frac{n}{2}} [{\cal P}_\Omega(\vec
\varphi_0')]^{\frac{n}{2}}}.\label{VV} \ee 
The last quantity is
directly calculable, since the Gaussian Ansatz asserts that 
\be
\overline {\prod_{a=1}^n {\cal P}_\Omega(\vec \varphi_0^a)} =
\sum_{\pi} \exp [- \frac{\beta}{2} \vec \varphi_0^{\pi(a)}
G^{-1}_{ab}({{\bf k}}_0)\vec \varphi_0^{\pi(b)}],\label{Perm} \ee
where $G$ is the optimal matrix determined {\it via} the
self-consistent equations and $\pi$ denotes all the permutations of
the replica indices. (All the saddle points only differing by
permutation of the indices must be taken into account). The quantity
in the right hand side of Eq. (\ref{VV}) corresponds to the choice
$\varphi_0^a=\varphi_0$ for $\frac{n}{2}$ indices, and
$\varphi_0^a=\varphi_0'$ for the other $\frac{n}{2}$. The next trick
to compute (\ref{Perm}) is to notice that in this case one can write
$\varphi_0^a \equiv {1\over 2} \left[\varphi_0 (1+\sigma_a) +
\varphi_0'(1-\sigma_a) \right]$, where $\sigma_a = \pm 1$ are
fictitious Ising spins which pick up a particular permutation,
provided $\sum_{a=1}^n \sigma_a = 0$.  The technique for working out
the sums over such spin configurations has been developped in the
appendix (D) of ref.\cite{BMP}.  Within a Parisi ansatz for the matrix
$G$, the final result for the force correlation, written in the case of $N=1$ to keep notations simple, is the following:

\be
\overline {f_\Omega( \varphi_0)f_\Omega( \varphi_0')}= \frac{2}{\beta} ck_0^2 -
\frac{4}{
\beta^2} \frac{\partial^2}{\partial \varphi_0 \partial
\varphi_0'} \int_0^\infty dh \Psi(h,u=0) \label{fcorr}
\ee
where $\Psi(h,u)$ satisfies
a non linear partial differential equation: 
\be -\frac{\partial
\Psi}{\partial u} = \frac{1}{2}\frac{dq}{du} \big( \frac{\partial^2
\Psi}{\partial h^2} + u [\frac{\partial\Psi}{\partial h}]^2 \big),
\label{edp} \ee 
where $0 \leq u \leq 1$ is the Parisi variable, indexing the pairs
of replica indices $a \neq b$ in the limit $n \to 0$.  The function
$q(u)$ is related to the matrix $G({{\bf k}},u)$ through: 
\be
 q(u) = -
\beta \frac{(\varphi_0-\varphi_0')^2}{4} G^{-1}({{\bf k}}_0,u) \label{qu}
\ee
and
the boundary condition is
\be \Psi(h,u=1)=
\ln\((1+\exp[-2h-2q(1)+2\int_0^1 du \ q(u)]\)) \ .  \ee 
Hence, once
$G({{\bf k}},u)$ is determined, the correlation function of the
effective potential acting on mode ${{\bf k}}_0$ is determined by
solving (\ref{edp}), which depends on $\varphi_0-\varphi_0'$ through
$q(u)$.

The solution of the self-consistent equations for $G({{\bf k}},u)$
was discussed in \cite{MP,BMY}. Let us specialize to the case $N =
\infty$, and introduce two important physical quantities, namely:

$\bullet$ The Larkin-Ovchinnikov length $\xi_{LO}$ separating a
`weakly distorted' regime for $|{{\bf x}}| < \xi_{LO}$, where all the
displacements induced by the random potential are small compared to
the correlation length of the potential $\Delta$, from a strongly
distorted regime. Simple dimensional arguments lead to \cite{LO,BMY}
\be \xi_{LO} \equiv \left(\frac{c^2 \Delta^4}{\hat
W}\right)^{\frac{1}{4-D}}, \ee 
where $\hat W$ is a rescaled potential
strength, defined as $\hat W= (2\pi)^2 W/(4-D)$. The reason for introducing
this rescaling in $\frac{1}{4-D}$ comes from the non trivial phase diagram around dimension $D=4$ \cite{BCDM}. Indeed, it is easily seen from the study
of the linearised, random force problem, that a `weak disorder' regime
with non trivial wandering exponent only exists when $\hat W$ is small
enough. If one keeps the original $W$ fixed and lets the dimension $D$
go to $4$, one enters a different phase (which actually survives for
$D>4$) \cite{BCDM}.

$\bullet$ A `Reynolds' number $Re$ (this terminology comes from the
analogy with Burgers' equation \cite{BMP}), defined as the ratio of
the elastic energy stored in a volume $\xi_{LO}^D$ to the temperature
$\frac{1}{
\beta}$.  We shall define $Re$ as
\be Re \equiv \beta {\cal
C}(D) \hat W^{\frac{2-D}{4-D}} (c\Delta^2)^{\frac{D}{4-D}}, \ee 
where
${\cal C}(D)$ is a dimension dependent number. Note that for
$D=4-\epsilon$ with $\epsilon$ small, ${\cal C} = \frac{\epsilon^2}{2}$.

We shall only consider the case of low temperature and weak disorder,
so that $Re \gg 1$ and $\xi_{LO} \gg a$, where $a =
\frac{2\pi}{\Lambda}$ is the small scale lattice constant which
regularizes the integrals over ${{\bf k}}$.  Under these conditions,
we obtain the following result for $D>2$: \ba G^{-1}({{\bf k}},u) &=&
- \sigma_1 \left(\frac{u}{u_c}\right)^{\frac{4-D}{D-2}} \qquad
\mbox{for} \quad u \leq u_c \\ &=& - \sigma_1 \qquad \qquad \qquad
\mbox{for} \quad u_c \leq u \leq 1 \ea with $\sigma_1=\frac{
\beta \hat
W \epsilon}{\Delta^2}$ and $u_c = \frac{1}{Re}$. 

The non trivial dependence of $G^{-1}({{\bf k}},u)$ on $u$
corresponds to continuous `replica symmetry breaking'. Let us now
analyze the partial differential equation (\ref{edp}) in the limit
$\beta \to \infty$. To this aim, we introduce the
notation $\gamma=\frac{6-2D}{D-2}$ and the following rescaled
variables: 
\be u = u_c v \qquad \psi = u_c\Psi \qquad g=\beta \sigma_1
u_c^2 \left(\frac{4-D}{8(D-2)}\right) (\vec \varphi_0-\vec
\varphi_0')^2 \qquad z=u_c h/ \sqrt{g}. \ee 
Eq. (\ref{edp}) then
transforms into 
\be \frac{\partial \psi}{\partial v} = - v^\gamma
[\psi'' + v \psi'^2],\label{edpr} \ee 
(the $'$ means
$\frac{\partial}{\partial z}$), with boundary condition (in the limit
$\beta \to \infty$): 
\be \psi(z,v=1)=-2\(( (D-2) g + z \sqrt{g}\))
\Theta\left(-(D-2) g - z \sqrt{g}\right),\label{Init} \ee 
where
$\Theta$ is the step function. The correlation of free energies (\ref{fcorr}) 
thus involves,
after change of variables, the integral
\be {\cal I}\equiv \int_0^\infty dh \Psi(h,u=0) = - 
\frac{\sqrt{g}}{u_c^2} \int_0^\infty dz \ z \psi'(z,v=0).\label{Idef} \ee 
Under
this form, the problem of evaluating $\cal I$ for small $|\vec
\varphi_0-\vec \varphi_0'|$ can simply be treated by solving
Eq. (\ref{edpr}) for $\psi'(z,v)$, perturbatively in $g$ -- see Appendix A. The result
reads: 
\be {\cal I} = \frac{1}{u_c^2} \left[\frac{g}{\gamma+1} - {\cal
L}(\gamma) g^{3/2} \right], \ee 
with ${\cal L}(\gamma)$ a complicated function of $\gamma$.  In the limit
$D=4-\epsilon$, $\gamma \simeq -1+\frac{\epsilon}{2}$, and ${\cal
L}(\gamma) \simeq 2\sqrt{\pi}$. Transforming back to the
original variables, we find, in the limit $k_0 \to 0$ : 
\be
R_{RSB}(y)-R_{RSB}(0)=- \epsilon \hat W \frac{y}{2\Delta^2} \(( 1 - \frac{\sqrt{\pi}} {2} \frac{\sqrt{y}} {\Delta \xi_{LO}^{D/2}} \)) ,
\label{RRSB}
\ee 
with $y=(\vec \varphi_0-\vec \varphi_0')^2$. Quite remarkably, Eq. (\ref{RRSB}) has the same form as the FRG result, Eq. (\ref{R*}), provided $\hat W$ is chosen in such a way that $\xi_{LO}$ remains fixed as $D \to 4$. This
$y^{\frac{3}{2}}$ behaviour was first obtained within a replica theory in
\cite{BMP} in the case $D=1$ (corresponding to Burgers' turbulence),
where the solution has a simpler, `one-step' structure (valid for
$D<2$): $G^{-1}({\bf k},u)=-\sigma_1 \Theta(u-u_c)$.

\section{Physical interpretation}

\subsection{Shocks and relationship with the Burgers' equation}

As mentioned above, the Gaussian variational ansatz does not mean
that the statistics of $V^*_\Omega$ is Gaussian. Let us first discuss
the replica construction of the effective potential in the simpler
case $D=1$ where a one step solution holds \cite{MP,BMP}. In this
case, one has: 
\be V^*_\Omega(\varphi)=-\frac{1}{\beta} \ln
\left[\sum_\alpha e^{-\beta F_\alpha -\frac{(\varphi -
\varphi_\alpha)^2}{u_c \Delta^2}}\right], \label{valpha} \ee 
where
$\alpha$ label the `states', centered around $\varphi_\alpha$ and of
free-energy $F_\alpha$, both depending on the `sample' $\Omega$. The
major prediction of the replica theory is that the $F_\alpha$ are
exponentially distributed for `deep' states {\footnote{This can
actually be understood within the general context of extreme event
statistics. \cite{BM96}}}, i.e: 
\be \rho(F_\alpha) \propto_{F_\alpha
\to -\infty} \exp(-\beta u_c |F_\alpha|). \label{expq} \ee 
The full
distribution of the effective force $\partial V^*_\Omega \over
\partial \vec \varphi_0$ (corresponding to the velocity in the Burgers
problem) was analyzed in detail in \cite{BMP}. Using the turbulence
language, it was found that the velocity field organizes in a
`froth-like' structure of $N-1$ dimensional shocks of vanishing width
in the limit $Re \to \infty$. Correspondingly, the potential has for $N=1$ the
shape drawn in Fig. 1: it is made of parabolas matching at angular
points -- the shocks. The singular behaviour of the force-force
correlation function, Eq. (\ref{f*}), is due to the fact that with a
probability proportional to the `distance' $|\vec \varphi_0 -\vec
\varphi_0'|$, there is a shock which gives a {\it finite} contribution
to $\vec f(\vec \varphi_0)-\vec f(\vec \varphi_0')$. This means in
particular that all the moments $\overline{|\vec f(\vec
\varphi_0)-\vec f(\vec \varphi_0')|^p}$ grow as $|\vec \varphi_0-\vec
\varphi_0'|$ for $p \geq 1$, instead of $|\vec \varphi_0-\vec
\varphi_0'|^{\frac{p}{2}}$ as for Gaussian statistics. It is not clear how this strong
departure from Gaussian statistics can be incorporated in an FRG treatment (see Section 4.C
below).

The relation with Burgers' equation is not coincidental and actually
quite interesting. Keeping $N=1$ for simplicity, consider a toy model
for the FRG mode elimination in which the renormalized effective
potential is defined as
\be \beta V_R(\varphi_<) = - \ln \left[\int d\varphi_>
e^{-\beta [\frac{c\Lambda^2}{2} \varphi_>^2 + V_0(\varphi_< +
\varphi_>)]} \right].\label{toy} \ee 
This means that $V_R(\varphi_<)$ is
precisely the Cole-Hopf solution of the Burgers equation
\cite{Burgers}: 
\be \frac{\partial V(\varphi,t)}{\partial t} =
\frac{1}{2\beta c \Lambda^2} \frac{\partial^2 V(\varphi,t)}{\partial
\varphi^2} - \frac{c \Lambda^2}{2} \left(\frac{\partial
V(\varphi,t)}{\partial \varphi}\right)^2 \ee
with 
\be
V(\varphi,t=0)=V_0(\varphi) \qquad V_R(\varphi)=V(\varphi,t=1). \ee 
As
is well known \cite{Burgers,Kida}, a random set of initial conditions
(here the bare pinning potential acting on $\varphi$) develops shocks
which separates as time grows, between which the `potential'
$V(\varphi)$ has a parabolic shape. Elimination of fast modes in a
disordered system thus naturally generates a `scalloped' potential,
with singular points (which are smoothed out at finite temperature or
finite $Re$) separating potential wells -- the famous `states'
appearing in the replica theory. Quite remarkably, this structure was
anticipated in \cite{Feigelman,DFI} using different arguments.

\subsection{Full RSB and multiscale effective potential}

In the case of continuous RSB, the effective potential is recursively
constructed via a set of `Matrioshka doll' Gaussians. It is
schematically drawn in Fig 2 for the the transverse fluctuations 
$\phi(\ell) - \phi(0)$. For each length scale $\ell$, one can define a characteristic value of the parameter $u(\ell)$ which plays the role of $u_c$ in Eq. (\ref{expq}) and sets the scale of the energy fluctuations. $u(\ell)$ is 
such that the diagonal part of
$G^{-1}({k}_0=\frac{2\pi}{\ell})$, namely $c {k}_0^2$, is equal to the off diagonal part
$G^{-1}({k}_0,u)$, which gives $u(\ell) \propto
\frac{1}{\beta}\((\frac{\xi_{LO}}{\ell}\))^\theta$ ($\theta=D-2$ is the `energy'
exponent in the case $N = \infty$, and is related to the small $u$ power-law behaviour of $G^{-1}({k}_0,u)$). The large scale structure of the effective
potential is thus a succession of parabolas of depth $\propto
\ell^{\theta}$, but this envelope structure is decorated by
hierarchically imbedded parabolas corresponding to all the smaller
length scales, between $\ell$ and $\xi_{LO}$, beyond which the
shocks disappear, since one enters into the effectively replica
symmetric random force regime. The important point however is that
small scale shocks are much more numerous than large scale ones and
completely dominate the small $y$ behaviour of $R_{RSB}(y)$: see Fig
2. This explains why the above result (\ref{RRSB}) is independent of
$k_0$ and only reflects the structure of $G^{-1}(k,u)$ in the vicinity
of $u_c$, corresponding to $k \simeq \frac{1}{\xi_{LO}}$. On the
other hand, quantities like $\overline{[\phi(\ell)-\phi(0)]^2}$ are
dominated by the region where $u \simeq u(k_0={2 \pi \over \ell})$,
corresponding to large scale moves. More precisely, the main
contribution to $\overline{[\phi(\ell))-\phi(0)]^2}$ comes from minima
separated by a distance $\ell^\zeta$ which happen to be separated by
an energy gap smaller than the temperature \cite{MP,HF}. This occurs
with probability $\propto \beta^{-1}\times \((\beta u(k_0)\))$ (see
Eq. (\ref{expq})).

In other words, the effective potential calculated within the FRG
procedure involves an extra step which we have not performed within
the replica construction, which is a coarse graining of the $\phi$
variables. In the FRG calculation, one restricts to configurations
which are such that $\phi$ is constant on scales $\ell$, and scales as
$\ell^\zeta$ {\footnote{ Removing all the modes $k > k_0$ in the
replica calculation leads to a correlation function $R_{RSB}(y)$
indeed dominated by the vicinity of $u(k_0) \propto k_0^\theta$.}}. The
correct choice of $\zeta$ then ensures that there are only a few
shocks on the scale $\ell$. As we now discuss in a rather conjectural
way, this is perhaps why the FRG can still be controlled, the
departure from Gaussian statistics being in some sense `weak'.

\subsection{The FRG in the presence of shocks}

To understand the emergence of shocks in the FRG picture, and to
assess their impact on the perturbative procedure, it is useful to
study the above toy model for the renormalization group, defined by
Eq. (\ref{toy}), which amounts to discarding the internal degrees of
freedom.  Following Ref.~\cite{BF}, we write Eq. (\ref{toy}) at zero
temperature (and after a rescaling) as: 
\be V_{\rm R}(\phi_<) = \min_{\phi_>} \bigg\{ {1 \over
2}|\phi_>|^2 + V(\phi_<+\phi_>)\bigg\}. \label{toydef} \ee 
The
validity of the perturbative minimization scheme  was discussed
in detail in Ref.~\cite{BF}, {\sl assuming} Gaussian statistics for
the random potential $V$.  Errors occur in the perturbative minimization scheme  due to an incorrect
choice among multiple minima in the effective Hamiltonian for
$\phi_>$.  For a Gaussian potential, there is an extremely dense set
of such minima, and such an error occurs essentially with probability
one.  The FRG appears to be saved, however, because the {\sl
magnitude} of the resulting error in the energy is small (i.e. higher
order in $\epsilon$).

A rather different picture emerges if one assumes a smooth potential
with shocks (i.e. slope discontinuities in $V$) spaced by $O(1)$
distances. To understand the limitations of the perturbative minimization scheme  in this case,
consider the extremal condition of the toy model, \be \phi_> = - V'(
\phi_<+ \phi_>). \ee In a scalloped (piecewise quadratic) potential, a
perturbative solution in $ \phi_>$ converges to the minimal energy in
the local well containing $ \phi_>=0$.  For $|V|$ small , this is
indeed the global minimimum, {\it unless} a shock occurs within a
distance $|\phi_{\rm shock}| < O(|V|)$, as can be seen by examining
the effective Hamiltonian for $ \phi_>$ in the neighborhood of a cusp.
Provided that a shock is present, however, the incorrect minima is
chosen with a probability of $O(1)$, leading to a large error in
$V_{\rm R}$.  Thus for the scalloped potential, instead of persistant
small errors, the perturbative minimization scheme  is typically correct, but suffers from
catastrophic rare events that generate large errors with small
probability.

An interesting simplification occurs if one considers a periodic
random potential $V$.  Such periodic potentials occur in models of
pinned charge density waves\cite{FLR,NF} and random anisotropy XY
magnets\cite{FXY}.  It is straightforward to show that repeated
applications of the toy model iteration drive the potential towards a
form with a single symmetric cusp per period. {\footnote{This is because a potential with multiple cusps
generically has multiple wells with non-uniformly separated and
unequal minima, so that repeated minimization allows the surviving
modes to remain only in the deepest well.}} For
such a symmetric form, the perturbative minimization scheme  {\sl always} converges to the correct
(deepest) minima of the effective potential, i.e. the local minimum is
always the global minimum.  Within the toy model, then, the perturbative minimization scheme 
appears to be {\sl asymptotically exact}.  Although errors may accrue
in early stages of the renormalization, these decrease as the length
scale grows and the final fixed point form is exact -- provided the
perturbation theory is carried out to all orders, of course! That the
FRG and replica methods lead to essentially the same results in this
case was underlined in \cite{LDG}.

The FRG consists, as does any renormalization group, of two parts: the
mode elimination (accomplished via the perturbative minimization scheme) and the rescaling
transformation.  The toy model allows a detailed study, in a somewhat
schematic way, of the former.  Within this framework, the
non-analyticity of $R$ emerges in a natural way via the generation of
Burgers' shocks.  The toy model, however, completely neglects the
internal degrees of freedom of the manifold, whose rescaling is
crucial for the power-counting in the full FRG.  In particular, this
rescaling not only leads to the existence of a fixed point for
$R(\phi)$, but also formally renders the higher cumulants of $V$
strongly irrelevant.

There appears to be a degree of competition between the mode
elimination, which favors shocks and the corresponding highly
non-Gaussian distribution for $V$, and the coarse graining and
rescaling transformation, which tends to keep the density of shocks to
a low value (at least for small $\epsilon$).  A complete description,
which is unfortunately not available to us at present, should properly
balance these effects against one another.  The special considerations
applicable for the periodic potential discussed above suggest that the
FRG may indeed be well controlled in that case.  More generally, the
full accommodation of shocks into the FRG remains a challenging open
problem.

\subsection{The $1+1$ Directed Polymer}

An explicit model where this construction actually does not require
the use of replicas or of the FRG is the $N=1$, $D=1$ (Directed
Polymer) case. From independent arguments \cite{HHF,HH}, one knows
that the effective potential $V_{x}(\phi)$ acting on the `head' of an
infinitely long polymer ($x \to \infty$) is a `random walk' in $\phi$
space: $\overline{[V_{x}(\phi)- V_{x}(\phi')]^2} \propto |\phi -
\phi'|$. (Notice the difference with Eq. (\ref{f*}), which concerns
the {\it force}, and not the potential). In particular, there are no
shocks in $V_{x}(\phi)$. Shocks appear when one coarse-grains the
description on a scale $\delta$. Let us define a coarse-grained
potential on an infinitesimal scale $\eta$ as
\be V_x^{\eta}(\rho)
\equiv -\frac{1}{\beta} \ln \int_{-\infty}^\infty d\phi \quad {\cal
K}\((\frac{(\phi-\rho)^2}{2 \eta}\)) e^{- \beta V_x(\phi)}, \label{1d}
\ee 
where ${\cal K}$ is an arbitrary local `filter'.  Iterating this
procedure a large number of times $\frac{\delta}{\eta}$ produces an
effective potential $V_x^\delta$ which, again, satisfies a Burgers
equation, but now with a long range correlated `initial condition'
$V_x(\phi)$. As is well known \cite{Burgers,Kida}, shocks also appear
in this case, with an average spacing growing as $\delta^{2/3}$. The
distribution of distance $d$ between shocks furthermore diverges for
small $d$ as $d^{-1/2}$ \cite{Kida}, indicating that there are shocks
on all scales smaller than $\delta^{2/3}$. All these results can
alternatively be obtained within the replica framework
\cite{Toy,BM96}.

\section{Discussion and Perspectives}

We have shown in this paper that the FRG and RSB techniques are not
contradictory but complementary. They both suggest quite an
appealing physical picture: the phase-space of the system is, on large
length scales, divided into `cells' corresponding to favourable
configurations where the potential is locally parabolic, and whose
depth is exponentially distributed. These cells are themselves
subdivided into smaller cells, corresponding to larger length scales,
etc.. This hierarchical construction is similar to the one usually
advocated for the phase space of spin-glasses \cite{exp}, based on
Parisi's RSB solution of the SK model \cite{MP,DFI}. The enormous
advantage of random manifolds is that this construction can be
directly performed in physical space.

An important consequence of this construction is that it allows us to
discuss the dynamical properties for
finite $N$ {\footnote{Infinite $N$ properties have been thoroughly
investigated in \cite{MFaging} but might belong to a different dynamical
class since activated effects disappear, at least for short range correlations}}. In the case of a one-step RSB, one can directly calculate
from Eq. (\ref{valpha}), the distribution of the height of the {\it
barriers} $\Delta E$ between two neighbouring wells, and finds that it
decays exponentially as $\exp(-
\beta u_c \Delta E)$. It is interesting
to notice that the barriers thus behave in the same way as energy
depths \cite{IV}, a point recently studied in detail for randomly
pinned lines in \cite{Drossel}. A natural picture for the dynamics is
thus to imagine that the manifolds jumps from well to well, each of
which representing a long-lived conformation of the manifold. Such a
picture is corroborated by recent numerical simulations in $D=1$,
$N=1$ \cite{Yoshino}. The lifetime of each `trap' is activated $\tau
\simeq \tau_0 \exp(\beta \Delta E)$, and is thus distributed as a
power-law $\tau^{-1-u(k)}$ for large $\tau$, where the exponent $u(k)
\propto k^\theta$ depends on the `size' of the jump (i.e. the mode
involved in the change of conformation), small $u(k)$ corresponding to
large wavelengths. Then, as emphasized in \cite{Bou92} where precisely
the same `trap' picture was advocated for spin-glasses, the dynamics
becomes non stationary and aging effects appear at low temperatures
and/or long-wavelengths such that $u(k) < 1$. For example, the
response of the manifold to a spatially modulated external field is
expected to behave, for $t \ll t_w$, as $(\frac{t}{t_w})^{1-u(k)}$,
where $t_w$ is the time elapsed since the quench from high
temperature. Correspondingly, the a.c. response should behave, for
$\omega t_w \gg 1$, as $(\omega t_w)^{u(k)-1}$, again much in the same
way as observed in spin-glasses \cite{Bou92}. For finite $N$ however,
one may expect that the exponential distribution of deep states
ceases to be valid outside the scaling region, i.e. for $\Delta E >>>
\frac{1}{\beta u(k)}$ \cite{Bou92,BM96}. This will lead to
`interrupted aging' for modes such that $\ln t_w >>> u(k)^{-1}$. These
equilibrated modes thereafter only contribute to the stationnary part
of the response (or correlation).

It is thus rather satisfactory that the `traps' appear naturally in
the context of pinned manifolds through the replica description, and
that this picture actually complement the `droplet' construction. It
would of course be gratifying to understand precisely how these ideas
could be extended to finite dimensional spin-glasses.

\vskip 2 true cm

\noindent {\it Acknowledgments} We would like to thank 
 E. Br\'ezin, C. De Dominicis and M. Feigel'man for 
interesting discussions.  L.B.'s work has been supported by the National Science Foundation under grant No.  PHY94--07194. M.M. thanks the Service de Physique Th\'eorique (Saclay) for its hospitality.

\appendix
\section {Perturbation expansion for Eq. (27)}

We provide here some intermediate steps of the computation of
free energy correlations with the replica method. We need to
solve eq.(\ref{edpr}) with the boundary condition (\ref{Init}),
and compute then the integral ${\cal I}$ defined in (\ref{Idef}).
The limit of interest is $g$ small. We work with the derivative
$\chi(z,v)=\psi'(z,v)$ which satisfies the equation:
\be
 \frac{\partial \chi}{\partial v} = - v^\gamma
[\chi'' + v \chi \chi']
\ee
together with the boundary condition:
\be
 \chi(z,v=1)=-2  \sqrt{g}
\Theta\left(- (D-2) g - z \sqrt{g}\right) \ .
\ee
The solution to this differential equation to order $g$ can be written as:
\be
\chi(z,v=0)=\chi_1(z)+\chi_2(z)
\ee
where
\be
\chi_1(z)=-2 \sqrt{g} \int_{-\infty}^{-(D-2)\sqrt{g}} \frac{dz'}{\sqrt{\frac{4 \pi}{\gamma+1}}} \exp-\((\frac{(z-z')^2}{{4 \over \gamma+1}}\))
\ee
and
\be
\chi_1(z)= 4 g \int_0^1 v^{\gamma+1} dv \int_{-\infty}^{\infty} \frac{dz'}{\sqrt{\frac{4 \pi v^{\gamma+1}}{\gamma+1}}} \exp-\((\frac{(z-z')^2}{{4v^{\gamma+1}\over \gamma+1}}\)) \frac{\partial}{\partial z'}
\((\int_{-\infty}^0  \frac{dz_1}{\sqrt{\frac{4 \pi (1-v^{\gamma+1})}{\gamma+1}}} \exp-\((\frac{(z-z')^2}{{4(1-v^{\gamma+1})\over \gamma+1}}\))
\))^2
\ee
Introducing the notation
\be
{\cal M}_0(x) = \int_x^\infty \frac{du}{\sqrt{2\pi}} e^{-u^2/2}
\ee
we find, after multiplication of $\chi$ by $z$ and integration:
\be
{\cal I} = \frac{1}{u_c^2} \(( \frac{g}{\gamma+1} - \sqrt{\frac{2}{\gamma+1}}
g^{3/2} [\sqrt{\frac{2}{\pi}}(D-2) - 4 \int_0^1 v^{\gamma+1} dv \int_{-\infty}^{\infty} dx {\cal M}_0(-\frac{x}{v^{\gamma+1\over 2}})
{\cal M}_0^2 (\frac{x}{\sqrt{1-v^{\gamma+1}}}) ] + ...\))
\ee
Expansion of the last integral for $\epsilon=4-D$ small, with $\gamma=-1+\epsilon/2$, reveals that the coefficient of $g^{3/2}$, which to leading order should be $\propto \epsilon^{-1/2}$ in fact vanishes, the next term being of order $\epsilon^0$.

\vskip 1 true cm

{\bf Figure Captions}

Fig. 1. Schematic view of the effective energy landscape as a
succession of parabolic wells matching at singular point. This picture
actually corresponds to a `one-step' replica symmetry breaking scheme.

Fig 2. Multiscale energy landscape corresponding to a full replica
symmetry breaking scheme. In this case, the construction is that of
parabolas within parabolas, in a hierarchical manner. The depth of the
wells (and thus also the height of the barriers) typically grows as
$|\phi - \phi'|^{\theta \over \zeta}$. The figure actually corresponds
to a two-step breaking scheme, with $u_1=0.5$ and $u_0=0.05$. The
inset is a zoom on a particular region, showing the first level of
Gaussians.

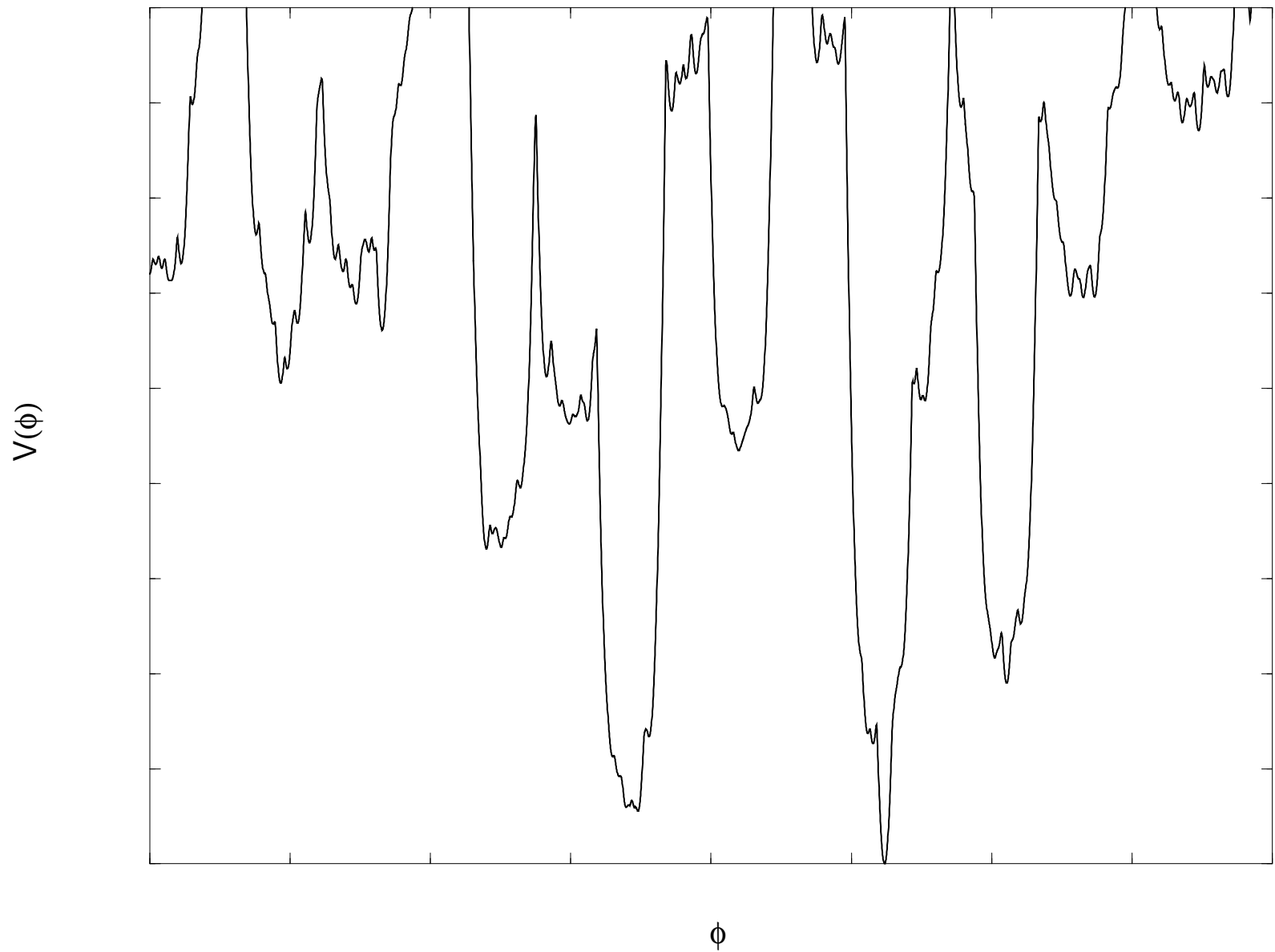
\end{document}